# Geometric quantification of photonic 4D spin-orbit states

Liang Fang[1]*†, Jinman Chen[1]†, Jia Cheng[1], Xuqi Guo[1], Senlin Huang[1], Qinjun Chen[1], Chujun Zhao[1], Shuangchun Wen[1], and Jian Wang[2]

[1]School of Physics and Electronics, Hunan University, Changsha 410082, Hunan, China.
[2]Wuhan National Laboratory for Optoelectronics, Huazhong University of Science and Technology, Wuhan, 430074, Hubei, China.
*Corresponding author: liangfang@hnu.edu.cn
†These authors contributed equally

High-dimensional photonic states have significantly advanced the fundamentals and applications of light. However, it remains huge challenges to quantify arbitrary states in high-dimensional Hilbert spaces with spin and orbital angular momentum bases. Here we introduce a geometric method to quantify arbitrary states in a 4D Hilbert space by interferometrically mapping them to unified centroid ellipses. Specifically, nine Stokes parameters can be deduced from three ellipses to quantify the 4D spin-orbit states described by SU(4) Poincaré hypersphere. We verify its feasibility by detecting these spin-orbit states gotten by both free-space wave plates and few-mode fibers. For the first time, we completely quantify and reconstruct higher-order modal group evolution of a weakly guiding few-mode fiber under twist perturbation. This geometric quantification, beyond the classical Stokes polarimetry, may pave the way to multi-dimensional optical metrology, sensing, and high-dimensional classical or quantum communications.

Light contains the fundamental amplitude (power), phase, wavelength (frequency), and polarization degrees of freedom (DoFs). Accordingly, the detection tools involve the classical photoelectric detector, interferometer, optical spectrometer, and Stokes polarimeter, recently developed into miniaturized and multi-functionalized devices[1-4]. Beyond these DoFs, light can be expanded in infinite spatial dimensionality, and thus be upgraded to a big family of structured light (Fig. 1a). This emerging structured light plays important roles in optical manipulation, high-resolution image, microscopy, multi-dimensional metrology, and classical or quantum information processing and communications[5-22]. Typically, as the solutions of paraxial wave equation in cylindrical coordinates, the higher-order Laguerre Gaussian (LG) beams carry intrinsic orbital angular momentum (OAM) with $\ell\hbar$ per photon, where the integer $\ell$ stands for the topological charge, and $\hbar$ is the Planck's constant. Such OAM allows us to infinitely expand the Hilbert space [5,6,16], beyond the binary spin angular momentum (SAM) with $\pm\hbar$ per photon. In mathematical representation, arbitrary superposition states with conjugate OAM or/and SAM bases can be completely described on Poincaré spheres or hypersphere[16,23-25].

The classical Stokes polarimetry enables complete quantification of light beams with spatially uniform polarization states as SAM superpositions[26,27]. Specially, when used for characterizing thin films and surfaces, it is commonly known as polarization-based ellipsometry. For the spatially inhomogeneous structured beams, up to now, the existing approaches to detecting or reconstructing them refer to photocurrent detection[28,29], photonic integrated processor[30,31], and mode decomposition with iterative algorithm or metasurfaces[32-35]. However, it still remains a huge challenge to completely quantify the arbitrary superposition states on high-dimensional Poincaré spheres, because of the requirement of complex polarization and modal decomposition, let alone the arbitrary four-dimensional (4D) spin-orbit states (Figs. 1b, 1c, and S1). Particularly, the latter are the common states as the degenerated higher-order modes of cylindrically symmetric waveguides, for example, the familiar few-mode fibers (FMFs) under perturbation[36,37].

The waveguiding higher-order modes can be analogous to higher-level bound states as the solutions of Schrödinger equation with a deep potential well, while the refractive index contrast plays the same role of the potential energ[38-40]. The cylindrically symmetric modes or states in highly confined conditions undergo the spin-orbit interaction that gives the out-of-step propagation between the conjugate spin or orbital components, associated with the degenerated modes lifting into vector eigenmodes (Fig. 3b) [41]. This belongs to an intrinsic spin-orbit perturbation that may produce the propagation-dependent polarization or mode pattern rotation[42,43]. Besides, the higher-order modes in the fiber systems may be easily perturbed from many extrinsic factors, such as various birefringences given by anisotropic temperature, stress, and twist on fibers. Especially, twist perturbation can lead to circular polarization or/and modal birefringences due to the SAM- or/and OAM-related Berry phase[44,45]. Absolutely, it is of great importance for optical communications and sensing to ascertain the higher-order mode evolution under these perturbations, which can be intuitively shown on Poincaré spheres via nine Stokes parameters (see Fig. 1b).

**Centroid geometric mapping**

Here we introduce a geometric detection to achieve the complete quantification of 4D spin-orbit states (Fig. 1d), beyond the functions of traditional polarimetry and mode analyzers. The centroid positional and orbital parameters as useful detectable quantities, derived from structured interference patterns with broken rotational symmetry, were recently uncovered to possess the potentials of high-resolution metrology and OAM superposition quantificatio[46,47]. Definitely, it is predictable that more centroid orbital parameters could be exploited from orthogonally polarized interference patterns to quantify high-dimensional photonic states. Here the 4D spin-orbit states, described on Poincaré hypersphere using SU(4) algebra[16,48,49], can be generally parameterized by

$$|\Psi_3\rangle_\ell = \cos(\theta_3/2)|\Psi_2\rangle_\ell |\sigma_+\rangle + \sin(\theta_3/2)e^{i\varphi_3}|\Psi_1\rangle_\ell|\sigma_-\rangle, \quad (1)$$

and

$$|\Psi_j\rangle_\ell = \cos(\theta_j/2)|\ell\rangle + \sin(\theta_j/2)e^{i\varphi_j}|-\ell\rangle, (j = 1,2), \quad (2)$$

where $|\pm\ell\rangle = A_\ell(r)e^{i[\omega(-t+z/c)\pm\ell\phi]}$ represents two conjugate OAM bases in the cylindrical coordinate $(r, \phi, z)$. $A_\ell$ denotes the radial-dependent amplitude, $\omega$ is the angular frequency, $c$ is light speed, $e^{i\ell\phi}$ represents the OAM-dependent helical wavefront with a topological charge $\ell$, and $|\sigma_\pm\rangle = \mathbf{x} \pm i\mathbf{y}$ indicate two orthogonal circular polarization (SAM) bases.

Each state $|\Psi_m\rangle_\ell$ ($m = 1, 2, 3$) can be described on its respective Poincaré sphere with the coordinate parameters $\theta_m$ ($0 \leq \theta_m \leq \pi$) and $\varphi_m$ ($0 \leq \varphi_m \leq 2\pi$). These six DoFs determine the longitudes and latitudes on the surfaces of unit spheres, dependent on the relative amplitudes and phases between the bases at two poles, respectively. Here the scalar states $|\Psi_1\rangle_\ell$ and $|\Psi_2\rangle_\ell$ describe two conjugate OAM superpositions on modal Poincaré spheres[23,46]. It is the third one $|\Psi_3\rangle_\ell$ that characterizes the targeted 4D spin-orbit states on the third sphere, where the former two scalar states act as modal bases at its two poles incorporated with respective spin components, respectively (Figs. 1b and S1a). Actually, such representation is equivalent to the representation by reassigning modal bases at the north and south poles[16], for example, using vector modes $TM_{01}$, $TE_{01}$, even and odd $HE_{21}$.

In order to fully infer the superposition parameters of 4D spin-orbit states from the observable centroid information by coaxial interference, a suitable reference state should be given as $|\Psi_0\rangle = \sqrt{2}/2\,(|\ell+1\rangle|\sigma_+\rangle + |-\ell-1\rangle|\sigma_-\rangle)$, pointing to $(S_1, S_2, S_3) = (1,0,0)$ on a unit Poincaré sphere.

The $\sigma_-$ and $\sigma_+$ spin components of interference patterns between the targeted and reference states can be separated via a spin splitting setup (see Fig. 1d). These two components can be written as $|\Psi_S^p\rangle = |\Psi_3^p\rangle_\ell + \sqrt{2}/2\,|\Psi_0^p\rangle e^{ik\Delta z}$, where $k = \omega/c$, $p = L$ or $R$ indicates the $\sigma_-$ or $\sigma_+$ components, respectively, and $\Delta z$ is optical path difference. The centroid coordinates can be calculated as $(\bar{x}_p, \bar{y}_p) = \langle \Psi_S^p|\hat{r}|\Psi_S^p\rangle/\langle \Psi_S^p|\Psi_S^p\rangle$, where $\hat{r}$ is the position operator. Accordingly, for the $\sigma_-$ component,

$$\bar{x}_L = g_1[a_1\cos(k\Delta z + \varphi_3) + a_2\cos(k\Delta z + \varphi_1 + \varphi_3)], \quad (3)$$

$$\bar{y}_L = g_1[a_1\sin(k\Delta z + \varphi_3) - a_2\sin(k\Delta z + \varphi_1 + \varphi_3)], \quad (4)$$

where $g_1 = \sqrt{2}\int|A_\ell A_{\ell-1}|r^2 dr/\int[2\sin^2(\theta_3/2)|A_\ell|^2 + |A_{\ell-1}|^2]rdr$, $a_1 = \sin(\theta_3/2)\cos(\theta_1/2)$, and $a_2 = \sin(\theta_3/2)\sin(\theta_1/2)$. Similarly, for the $\sigma_+$ component,

$$\bar{x}_R = g_2[b_1\cos(k\Delta z) + b_2\cos(k\Delta z + \varphi_2)], \quad (5)$$

$$\bar{y}_R = g_2[b_1\sin(k\Delta z) - b_2\sin(k\Delta z + \varphi_2)], \quad (6)$$

where $g_2 = \sqrt{2}\int|A_\ell A_{\ell-1}|r^2 dr/\int[2\cos^2(\theta_3/2)|A_\ell|^2 + |A_{\ell-1}|^2]rdr$, $b_1 = \cos(\theta_3/2)\cos(\theta_2/2)$, and $b_2 = \cos(\theta_3/2)\sin(\theta_2/2)$ (see methods).

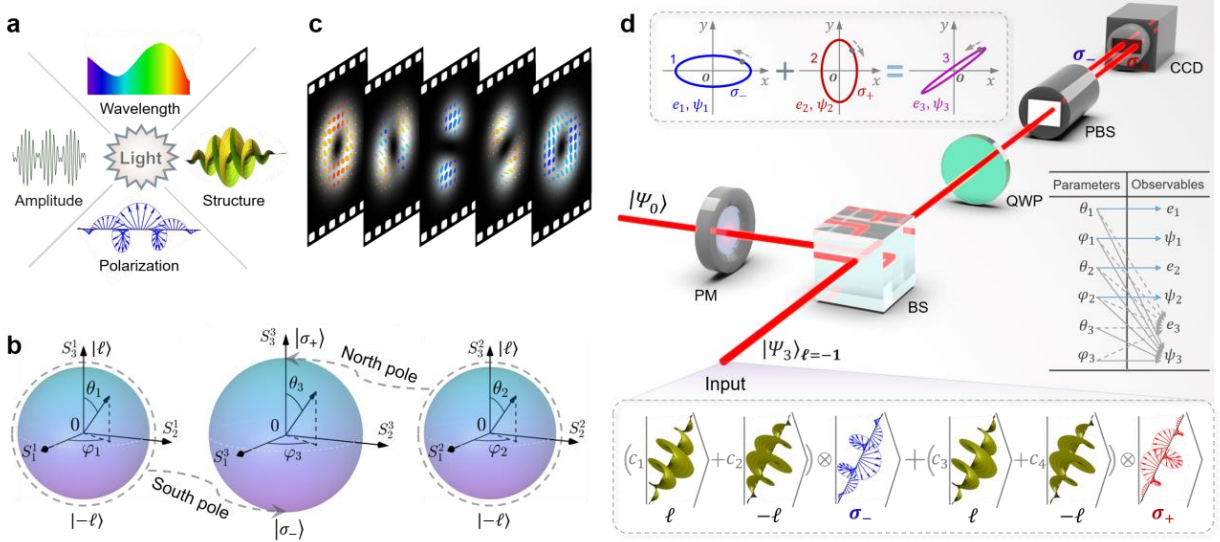

**Fig. 1 | Schematic of quantifying photonic 4D spin-orbit states by geometric mapping. a**, Electromagnetic DoFs including the classical complex amplitude, wavelength (frequency), polarization (spin), and the emerging spatial (OAM) structure. **b,** The 4D spin-orbit states as the superpositions between conjugate OAM and SAM states can be completely described on SU(4) hypersphere, where the modal bases at the north and south poles of middle spheres come from the arbitrary states marked with the numbers on the right and left spheres, respectively. **c**, Modal patterns of five 4D spin-orbit states with intensity and local polarization distributions. **d**, Key part of experimental setup. PM: phase modulator; BS: beam splitter; QWP: quarter wave plate; PBS: polarization beam splitter; CCD: charge coupled device. The input 4D spin-orbit states are characterized by arbitrary superpositions between two conjugate OAM (helical phase planes) and SAM (circular polarization states), where $c_1$, $c_2$, $c_3$, and $c_4$ are complex coefficients determined by six superposition parameters (DoFs). The table that shows geometric mapping from the six superposition parameters to six experimentally observable elliptic parameters. The upper inset shows that two centroid ellipses (red and blue ellipses) respectively extracted from orthogonal

interference patterns can superpose into the third correlative ellipse by taking mean values of the centroid coordinates.

In principle, all superposition parameters $\theta_m$ and $\varphi_m$ ($m = 1, 2, 3$) can be deduced from the two centroids in Eqs. (3)-(6) based on their relative positional information. Nonetheless, the data analysis and processing would be so complicated that it is hard to implement in practices. An applaudable method is to utilize the geometric information of elliptic orbits to retrieve the superposition parameters. Remarkably, both centroid coordinates ($\bar{x}_L, \bar{y}_L$) and ($\bar{x}_R, \bar{y}_R$) feature elliptical trajectories when changing optical path differences ($\Delta z$)[46], or modulating the phase of the reference path (Fig. 1d). The elliptic parameters in terms of ellipticities $e_{1,2}$ and orientation angles $\psi_{1,2}$ of major axes are mapped from the superposition parameters $\theta_{1,2}$ and $\varphi_{1,2}$, in return, one can get,

$$\theta_m = \pi/2 - 2\tan^{-1}(e_m), \tag{7}$$

and

$$\varphi_m = -2\psi_m, \tag{8}$$

where $m = 1, 2$. Furthermore, in order to retrieve the parameters $\theta_3$ and $\varphi_3$, another correlative orbit could be obtained by taking the mean values of centroid coordinates in Eqs. (3)-(6), given as ($\bar{x}_M, \bar{y}_M$) = $0.5(\bar{x}_L + \bar{x}_R, \bar{y}_L + \bar{y}_R)$, generally featuring an ellipse as well. Actually, to simplify mathematical solution and signal processing, the coefficients $g_1$ and $g_2$ can be controlled as a constant, being independent of the superposition parameters. This could be realized by reassigning half of the total optical power to both the denominators when algorithmically calculating the centroids, i.e., $\mathbf{r}' = \langle \Psi_S^p | \hat{r} | \Psi_S^p \rangle / \langle \Psi_S' | \Psi_S' \rangle$, where $|\Psi_S'\rangle = \sqrt{2}/2 \, (|\Psi_3\rangle_\ell + |\Psi_0\rangle)$. In this case, $g_1 = g_2 = g = \sqrt{2} \int |A_\ell A_{\ell-1}| r^2 dr / \int (|A_\ell|^2 + |A_{\ell-1}|^2) r dr$. From this correlative ellipse with $e_3$ and $\psi_3$, the parameters ($\theta_3, \varphi_3$) can be numerically calculated from

$$\tan(2\psi_3) = \frac{a_1 b_2 \sin(\varphi_3 - \varphi_2) - a_2 b_1 \sin(\varphi_3 + \varphi_1) - a_1 a_2 \sin(\varphi_1) - b_1 b_2 \sin(\varphi_2)}{a_1 b_2 \sin(\varphi_3 - \varphi_2) + a_2 b_1 \sin(\varphi_3 + \varphi_1) + a_1 a_2 \sin(\varphi_1) + b_2 \sin(\varphi_2)}, \tag{9}$$

$$\frac{2e_3}{1+e_3^2} = \frac{a_1^2 - a_2^2 + b_1^2 - b_2^2 + 2a_1 b_1 \cos(\varphi_3) - 2a_2 b_2 \cos(\varphi_3 - \varphi_2 + \varphi_1)}{1 + 2a_1 b_1 \cos(\varphi_3) + 2a_2 b_2 \cos(\varphi_3 - \varphi_2 + \varphi_1)}. \tag{10}$$

The derivation detail is presented in methods.

Thus, a useful geometric methodology is established to quantify the general 4D superposition states with parameters $\theta_m$ and $\varphi_m$ in Eq. (1) using the extractable six elliptic parameters $e_m$ and $\psi_m$, ($m = 1, 2, 3$) (Fig. 1d). These superposition parameters can be further transformed into nine Stokes parameters in Cartesian coordinates for global positioning on the SU(4) hypersphere. Note that in the Eqs. (7) and (10) above, the signs of ellipticities are determined by the orbiting (clockwise or counter clockwise) directions of time-varying centroids. In addition, some special cases reduced to 2D states, for example, the uniform polarization states or scalar fields, should be excluded, because they give rise to zero in the denominators. Nevertheless, actually these states could be more easily quantified by the simpler centroid positional or orbital information, but beyond the topic here.

**Experimental verification**

We first conduct the proof-of-concept experiment by generating and meanwhile detecting the 4D spin-orbit photonic states in free space. The 4D hybrid states under $\ell = -1$ in Eq. (1) were generated by the combination of half wave plate (HWP), QWP1, Q-plate, and QWP2 (see Figs. 2a and S3). When making interference between the 4D spin-orbit states and reference state, the optical path differences were produced by manually pressing the prism in one direction. Meanwhile, the dynamical interference patterns with orthogonal spin components were synchronously collected by a high-speed CCD (about 180 fps) (Figs. S3 and S4). Two centroid ellipses 1 and 2 with elliptic parameters were obtained by algorithmically fitting two groups of centroids extracted from interference patterns, while the correlative ellipses 3 were obtained by taking the mean values of these two. Especially, the plus or minus sign of the orbital ellipticity can be experimentally distinguished via the phase differences of peak frequencies in Fourier phase spectra between the x- and y-coordinate variation of centroid orbiting (Extended Data Fig. 1). In the experiment, two evolution paths of 4D spin-orbit hybrid states were produced by rotating the QWPs in front and back of the Q-plate, respectively (see Supplementary information , Sec. 3).

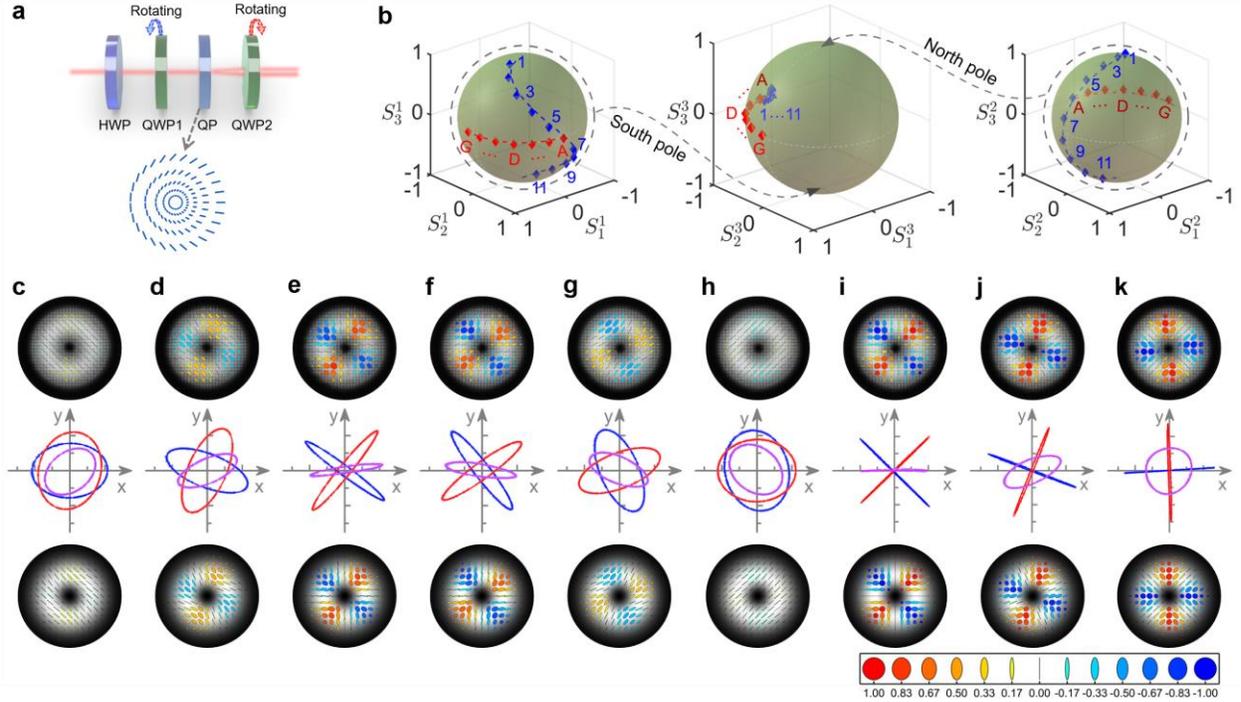

**Fig. 2│Experimental results of generating, quantifying, and reconstructing photonic 4D spin-orbit states by the proposed geometric method. a**, Free-space generation of 4D spin-orbit states by the combination of HWP, QWP1, Q-plate ($q = 1/2$), and QWP2. The inset shows the orientation angles of local fast axes of the Q-plate. **b**, Quantification results of the 4D spin-orbit states evolved along two paths on SU(4) hypersphere, produced by rotating QWPs in front (blue path) and back (red path) of the Q-plate shown in **a**, respectively. **c**-**k**, Measured results of these 4D spin-orbit states on the hypersphere marked by 1, 3, 5, 7, 9, 11, A, D, and G, respectively. The upper patterns show the produced modes that were reconstructed by the conventional Stokes decomposition. The middle figures show the centroid orbits (blue and red ellipses) extracted experimentally, and their correlative orbits (magenta ellipses). The down figures show the reconstructed patterns using LG beams with the parameters deduced from the three ellipses.

The quantification and reconstruction results along these two paths on SU(4) hypersphere are shown in Fig. 2b. In Figs. 2c-2k, the modal patterns (upper figures) of the produced 4D spin-orbit states were reconstructed using the conventional Stokes polarization decomposition[50], while the down modal patterns were reconstructed by using LG beams with the superposition parameters experimentally extracted from the three ellipses (middle figures). The local polarization states within patterns are indicated as ellipses with different ellipticities (shown in color-bar) and orientation angles of major axes. The comparison results between these reconstructed patterns using our centroid geometric methods and the conventional Stokes decomposition show good agreement. More measured patterns and comparison data are provided in Extended Data Figs. 2 and 3. Additionally, we offer a video show how the three orbital ellipses were extracted from two orthogonally polarized interference patterns (see Supplementary Video 1). The measured results by geometric detection here agree well with the predicted values on SU(4) hypersphere, as well as the reconstructed polarization patterns using the conventional Stokes decomposition. Note that the conventional method just enables the local polarization reconstruction of structured optical fields[50], but fails to quantify them on Poincaré spheres or hypersphere because of the unobtainable Stokes parameters (see Supplementary information Sec. 2 and Fig. S2). These results and comparisons show high feasibility and functionality of our geometric detection for high-dimensional photonic states.

**Monitoring Higher-Order Modes**

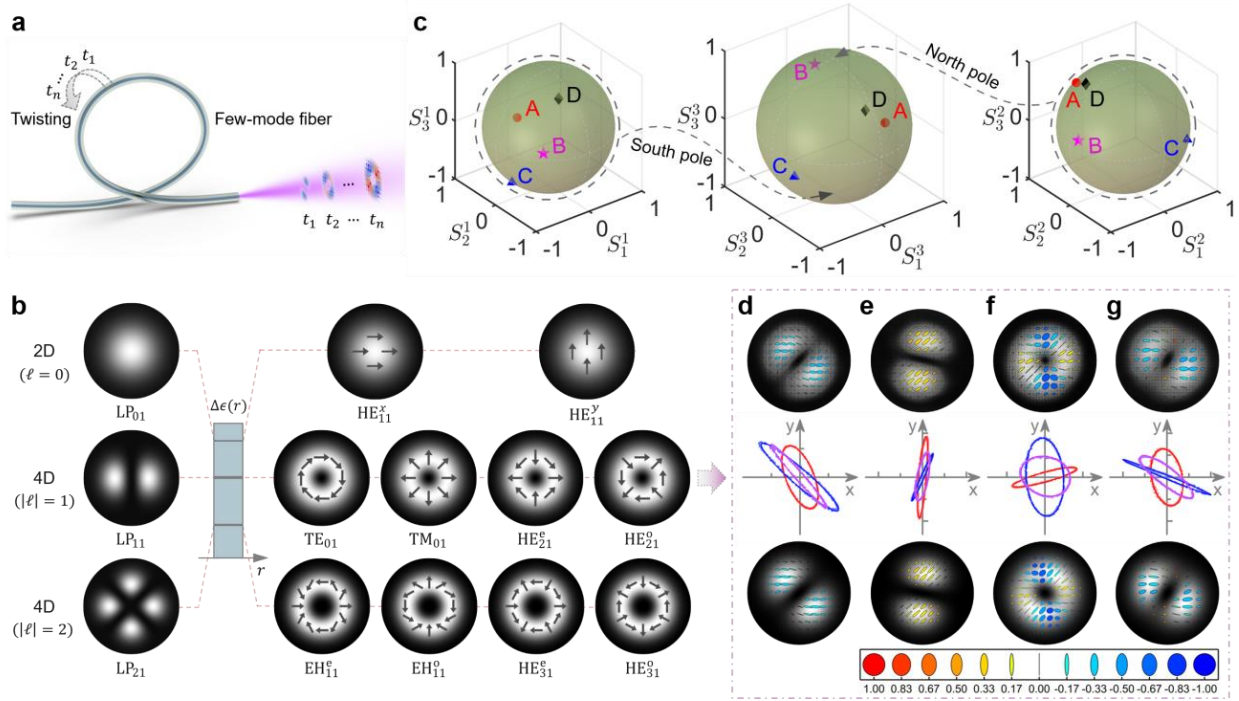

**Fig. 3 | Experimental demonstration of geometric quantification and reconstruction for arbitrary higher-order modes as 4D spin-orbit states outputted from a weakly guiding FMF. a,** Schematic of higher-order mode propagation and evolution by twisting the FMF. **b,** Classification and effective refractive indices of higher-order modes as the solutions of Maxwell's equations in cylindrically symmetric waveguides with permittivity distribution of $\Delta\epsilon(r)$. The left scalar fields under mode degeneracy contain the right 2D or 4D vector eigenmodes with different topological charges of OAM bases, where the arrows show the local polarization vectors. **c,** Experimental results of quantifying four 4D spin-orbit states (A, B, C, and D) on SU(4) hypersphere outputted from the FMF. **d-g,** Measured and comparison results of these four higher-order modes on the hypersphere, including the reconstructed modal patterns via the conventional Stokes decomposition (upper figures), the three ellipses experimentally extracted from two

orthogonally polarized interference patterns (middle figures), and the reconstructed patterns using LG beams with the superposition parameters deduced from the three ellipses (down figures).

As shown in Fig. 3, photonic 4D spin-orbit states can be naturally supported in cylindrically symmetric waveguide as the higher-order modal group under degeneracy. Besides the proof-of-concept measurement, we further apply our geometric detection to quantify and track the higher-order mode evolution outputted from a weakly guiding FMF. Here we excite the higher-order modes in the second modal group in the FMF and then twist it to make them evolution (see Figs. 3a and 3b). More experimental details are provided in Supplementary information . Firstly, four random 4D spin-orbit states were produced by manually twisting the FMF and then used for further verifying our geometric detection. The measured nine Stokes parameters based on the elliptic parameters experimentally extracted from interference patterns are positioned on the SU(4) hypersphere marked with different lowercases in Fig. 3c. Especially, we demonstrate the reconstructed patterns and their comparison results obtained by conventional Stokes polarization decomposition (Figs. 3d-3g). Obviously, the reconstructed modal patterns (down figures) based on the elliptic parameters (middle figures) are in good agreement with the measured patterns (upper figures) using the conventional method.

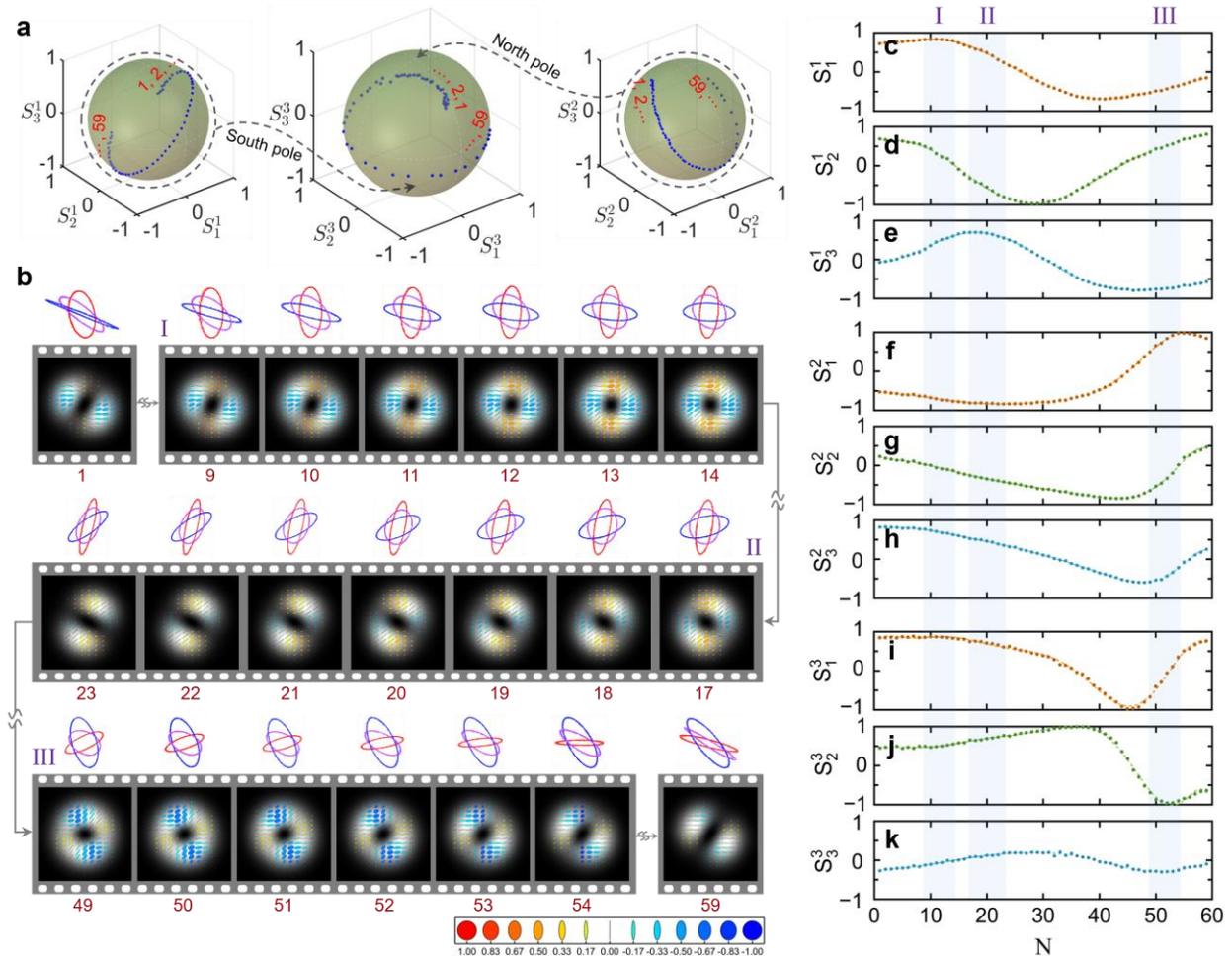

**Fig. 4 | Experimental demonstration of quantifying and reconstructing higher-order mode (4D spin-orbit states) evolution along a specific path produced by twisting the FMF. a**, The higher-order mode evolution is quantified on SU(4) hypersphere. **b**, Reconstructed higher-order

modal patterns in the evolution path based on the measured three ellipses shown above. Measured Stokes parameters of higher-order mode evolution on the left (**c-e**), right (**f-h**), and middle (**i-k**) spheres, respectively, where the details of the successive states I, II, and III, as well as beginning and end states, are shown in **b**. The dynamical demonstration and comparison results are demonstrated in Supplementary Video 2.

Furthermore, based on the principle of geometric detection, we perform continuous monitoring for the evolution of higher-order modes in the second group as typical 4D spin-orbit states under mode and polarization birefringences produced by stressing and twisting the FMF. The mode evolutions have been successfully quantified and monitored by the measured nine Stokes parameters on the SU(4) hypersphere (see Fig. 4). From the quantification results, remarkably, the slight changes in the successive states can be clearly distinguished through the parameter evolution of three ellipses (Fig. 4b). More details of the dynamical measured results for the two evolution paths are provided in Supplementary Videos 2 and 3. Note that in these videos we additionally present the intensity patterns directly collected by another CCD for comparison, because it has not enough time to perform the Stokes polarization decomposition with seven procedures during the process of geometric detection. We believe that this measurement of high-dimensional photonic states on SU(4) hypersphere here is a thorough upgrade for the conventional polarimetry that just enables the polarization detection on the classical Poincaré sphere. This may largely facilitate the applications of high-dimensional optical fields to optical communications and sensing.

**Summary and Outlook**

Photonic high-dimensional superposition states contain complex spatial phase and polarization information. Completely ascertaining and quantifying them is essential for high-dimensional classical or quantum information processing, multi-dimensional optical communications and fiber sensing[51-54]. It is yet far beyond the capabilities of the traditional optical intensity-based photoelectric detectors and polarimetry[55,56]. Here we introduce a full geometric mapping from photonic 4D spin-orbit states to three centroid ellipses through dynamical interference. The centroid orbital information shows a great potential to completely quantify the 4D photonic states by deducing the superposition parameters from the elliptic parameters. Its effectiveness has been substantiated by the rigorous mathematical derivation and the proof-of-concept experiments. Especially, based on this, for the first time, we successfully conduct the arbitrary monitoring of higher-order mode evolution of a commonly used weakly guiding FMF. Absolutely, this method is suitable for detecting higher-order modal states, for example, $\ell > 1$ in Eq. (1), or the third modal group in weakly guiding FMFs shown in Fig. 3b. This geometric detection may be hopefully developed into a useful characterization tool of high-dimensional structured light for multi-dimensional optical metrology, sensing, and beyond.

**References**

1. Yuan, S., Wang, Y., Zhang, C., Mei, T. & Qian, L. Geometric deep optical sensing. *Science* **379**, eade1220 (2023).
2. Rubin, N. A., D'Aversa, G., Chevalier, P., Shi, Z. & Chen, W. T. Matrix Fourier optics enables a compact full-Stokes polarization camera. *Science* **365**, eaax1839 (2019).
3. Yang, Z., Albrow-Owen, T., Cai, W. & Hasan, T. Miniaturization of optical spectrometers. *Science* **371**, eabe0722 (2021).
4. Fan, Y. et al. Dispersion-assisted high-dimensional photodetector. *Nature* **630**, 77-83 (2024).

5. Allen, L., Beijersbergen, M. W., Spreeuw, R. J. C. & Woerdman, J. P. Orbital angular momentum of light and the transformation of Laguerre-Gaussian laser modes. *Phys. Rev. A* **45**, 8185-8189 (1992).
6. Mair, A., Vaziri, A., Weihs, G. & Zeilinger, A. Entanglement of the orbital angular momentum states of photons. *Nature* **412**, 313-316 (2001).
7. Dorn, R., Quabis, S. & Leuchs, G. Sharper focus for a radially polarized light beam. *Phys. Rev. Lett.* **91**, 233901 (2003).
8. Marrucci, L., Manzo, C. & Paparo, D. Optical spin-to-orbital angular momentum conversion in inhomogeneous anisotropic media. *Phys. Rev. Lett.* **96**, 163905 (2006).
9. Willig, K. I. et al. STED microscopy reveals that synaptotagmin remains clustered after synaptic vesicle exocytosis. *Nature* **440**, 935-939 (2006).
10. Padgett, M. & Bowman, R. Tweezers with a twist. *Nat. Photonics* **5**, 343-348 (2011).
11. Fickler, R. et al. Quantum entanglement of high angular momenta. *Science* **338**, 640-643 (2012).
12. Shi, Z. et al. Super-resolution orbital angular momentum holography. Nat. Commun., 14, 1869 (2023).
13. Fang, L. et al. Vectorial Doppler metrology. *Nat. Commun.* **12**, 4186 (2021).
14. Cheng, M. et al. Metrology with a twist: probing and sensing with vortex light. *Light Sci. Appl.* **14**, 4 (2025).
15. Forbes, A., de Oliveira, M. & Dennis, M. R. Structured light. *Nat. Photonics* **15**, 253-262 (2021).
16. Zhang, Z. et al. Spin-orbit microlaser emitting in a four-dimensional Hilbert space. *Nature* **612**, 246-251 (2022).
17. Bliokh, K. Y. et al. Roadmap on structured waves. *J. Opt.* **25**, 103001 (2023).
18. Ma, Z. et al. Scaling information pathways in optical fibers by topological confinement. *Science* **380**, 278-282 (2023).
19. Bütow, J. et al. Generating free-space structured light with programmable integrated photonics. *Nat. Photonics* **18**, 243-249 (2024).
20. Zhan, Q. Spatiotemporal sculpturing of light: a tutorial. *Adv. Opt. Photon.* **16**, 163-228 (2024).
21. Shen, Y. et al. Optical skyrmions and other topological quasiparticles of light. *Nat. Photonics* **18**, 15-25 (2024).
22. Ye, G. et al. Recent progress on laser interferometry based on vortex beams: Status, challenges, and perspectives. *Opt. Laser Eng.* **172**, 107871 (2024).
23. Padgett, M. J. & Courtial, J. Poincaré-sphere equivalent for light beams containing orbital angular momentum. *Opt. Lett.* **24**, 430-432 (1999).
24. Milione, G. et al. Higher-order Poincaré sphere, Stokes parameters, and the angular momentum of light. *Phys. Rev. Lett.* **107**, 053601 (2011).
25. Gutiérrez-Cuevas, R. et al. Modal Majorana sphere and hidden symmetries of structured-Gaussian beams. *Phys. Rev. Lett.* **125**, 123903 (2020).
26. Martínez, A. Polarimetry enabled by nanophotonics. *Science* **362**, 751-752 (2018).
27. Deng, J. et al. An on-chip full-Stokes polarimeter based on optoelectronic polarization eigenvectors. *Nat. Electron.* **7**, 1004-1014 (2024).
28. Ji, Z. et al. Photocurrent detection of the orbital angular momentum of light. *Science* **368**, 763-767 (2020).
29. Gunyaga, A. A., Durnev, M. V. & Tarasenko, S. A. Photocurrents induced by structured light. *Phys. Rev. B* **108**, 115402 (2023).

30. Bütow, J. et al. Photonic integrated processor for structured light detection and distinction. *Commun. Phys.* **6**, 369 (2023).
31. Bütow, J. et al. Spatially resolving amplitude and phase of light with a reconfigurable photonic integrated circuit. *Optica* **9**, 939-946 (2022).
32. Shapira, O. et al. Complete modal decomposition for optical waveguides. *Phys. Rev. Lett.* **94**, 143902 (2005).
33. Manuylovich, E. S. et al. Fast mode decomposition in few-mode fibers. *Nat. Commun.* **11**, 5507 (2020).
34. Yang, H. et al. Metasurface higher-order Poincaré sphere polarization detection clock. *Light Sci. Appl.* **14**, 63 (2025).
35. Li, X. et al. Diffractive Metasurface-Enabled Single-Shot Characterization of Hybrid-Order Poincaré Beams. *Adv. Optical Mater.* **13**, 2403405 (2025).
36. Fang, L. et al. Spin-Orbit Mapping of Light. *Phys. Rev. Lett.* **127**, 233901 (2021).
37. Okamoto, K. *Fundamentals of Optical Waveguides* (Elsevier Academic, 2006).
38. Leary, C. C. & Smith, K. H. Unified dynamics of electrons and photons via Zitterbewegung and spin-orbit interaction. *Phys. Rev. A* **89**, 023831 (2014).
39. Bliokh, K. Y. et al. Semiclassical Dynamics of Electron Wave Packet States with Phase Vortices. *Phys. Rev. Lett.* **99**, 190404 (2007).
40. Guzmán-Silva, D. et al. Experimental Observation of Interorbital Coupling. *Phys. Rev. Lett.* **127**, 066601 (2021).
41. Fang, L. & Wang, J. From Imbert-Fedorov shift to topologically spin-dependent walking off for highly confining fiber-guided twisted light. *J. Opt.* **23**, 065603 (2021).
42. Vitullo, D. L. P. et al. Observation of Interaction of Spin and Intrinsic Orbital Angular Momentum of Light. *Phys. Rev. Lett.* **118**, 083601 (2017).
43. Leary, C. C., Raymer, M. G. & van Enk, S. J. Spin and orbital rotation of electrons and photons via spin-orbit interaction. *Phys. Rev. A* **80**, 061804 (2009).
44. Tomita, A. & Chiao, R. Y. Observation of Berry's Topological Phase by Use of an Optical Fiber. *Phys. Rev. Lett.* **57**, 937-940 (1986).
45. Gregg, P., Kristensen, P. & Ramachandran, S. Conservation of orbital angular momentum in air-core optical fibers. *Optica* **2**, 267-270 (2015).
46. Cheng, J. et al. Optical centroid ellipses beyond polarization ellipses. *Opt. Lett.* **50**, 97-100 (2025).
47. Fang, L. et al. Optical centroid orbiting metrology. *Laser Photon. Rev.* **19**, 2402311 (2025).
48. Kemp, C. J. D. et al. Nested-sphere description of the N-level Chern number and the generalized Bloch hypersphere. *Phys. Rev. Res.* **4**, 023120 (2022).
49. Sbaih, M. A. A. et al. Lie Algebra and Representation of SU(4). *Electron. J. Theor. Phys.* **10**, 9-26 (2013).
50. Liang, Y. et al. Reconfigurable structured light generation and its coupling to air-core fiber. *Adv. Photon. Nexus* **2**, 036015 (2023).
51. Ashry, I. et al. A review of using few-mode fibers for optical sensing. *IEEE Access* **8**, 179592-179605 (2020).
52. Leedumrongwatthanakun, S. et al. Programmable linear quantum networks with a multimode fiber. *Nat. Photonics* **14**, 139-142 (2020).
53. Cristiani, I. et al. Roadmap on multimode photonics. *J. Opt.* **24**, 083001 (2022).
54. Lamb, E. S. et al. Shape sensing endoscope fiber. *Optica* **11**, 1462-1468 (2024).
55. Goldstein, D. Polarized Light (CRC Press, 3rd edn, 2011).
56. Lissberger, P. H. Ellipsometry and polarised light. *Nature* **269**, 270-271 (1977).

## Methods

### Mapping photonic 4D spin-orbit states to multiple centroid elliptic parameters

From Eqs. (3)-(6), an important mapping relationship can be established from the superposition parameters $(\theta_m, \varphi_m)$ to the centroid elliptic parameters in terms of the ellipticity $(e_m)$ as the ratio of the semi-minor to semi-major axes, the orientation angle $(\psi_m)$ of the major axis, which can be expressed as

$$e_m = \tan\left(\frac{\pi}{4} - \frac{\theta_m}{2}\right), \tag{11}$$

$$\psi_m = -\frac{1}{2}\varphi_m, \tag{12}$$

where $m = 1$ and $2$ correspond to the $\boldsymbol{\sigma}_-$ or $\boldsymbol{\sigma}_+$ components, respectively. Furthermore, another useful correlative orbits can be obtained by taking the mean values of two centroids, given as,

$$\bar{x}_M = 0.5(\bar{x}_L + \bar{x}_R) = 0.5 g f_1 \cos(k\Delta z + \Delta\delta_1), \tag{13}$$

$$\bar{y}_M = 0.5(\bar{y}_L + \bar{y}_R) = 0.5 g f_2 \sin(k\Delta z + \Delta\delta_2), \tag{14}$$

where the new variables are defined as $f_1 = \sqrt{A_1^2 + A_2^2}$, $f_2 = \sqrt{B_1^2 + B_2^2}$, $\Delta\delta_1 = \tan^{-1}[A_2/A_1]$, and $\Delta\delta_2 = \tan^{-1}[B_2/B_1]$, and

$$A_1 = b_1 + a_2\cos(\varphi_3 + \varphi_1) + b_2\cos(\varphi_2) + a_1\cos(\varphi_3), \tag{15}$$

$$A_2 = a_2\sin(\varphi_3 + \varphi_1) + b_2\sin(\varphi_2) + a_1\sin(\varphi_3), \tag{16}$$

$$B_1 = b_1 - a_2\cos(\varphi_3 + \varphi_1) - b_2\cos(\varphi_2) + a_1\cos(\varphi_3), \tag{17}$$

$$B_2 = -a_2\sin(\varphi_3 + \varphi_1) - b_2\sin(\varphi_1) + a_1\sin(\varphi_3). \tag{18}$$

Thus, the mapping relationship for the middle sphere meets the equations

$$\tan(2\psi_3) = \frac{2f_1 f_2}{f_1^2 - f_2^2}\sin(\Delta\delta) = \frac{2(A_1 B_2 - A_2 B_1)}{A_1^2 + A_2^2 - B_1^2 - B_2^2} = \frac{a_1 b_2\sin(\varphi_3-\varphi_2)-a_2 b_1\sin(\varphi_3+\varphi_1)-a_1 a_2\sin(\varphi_1)-b_1 b_2\sin(\varphi_2)}{a_1 b_2\sin(\varphi_3-\varphi_2)+a_2 b_1\sin(\varphi_3+\varphi_1)+a_1 a_2\sin(\varphi_1)+b_1 b_2\sin(\varphi_2)}, \tag{19}$$

$$\sin(2\chi_3) = \frac{2e_3}{1+e_3^2} = \frac{2f_1 f_2}{f_1^2 + f_2^2}\cos(\Delta\delta) = \frac{2(A_1 B_1 + A_2 B_2)}{A_1^2 + A_2^2 + B_1^2 + B_2^2} = \frac{a_1^2 - a_2^2 + b_1^2 - b_2^2 + 2a_1 b_1\cos(\varphi_3) - 2a_2 b_2\cos(\varphi_3-\varphi_2+\varphi_1)}{1 + 2a_1 b_1\cos(\varphi_3) + 2a_2 b_2\cos(\varphi_3-\varphi_2+\varphi_1)}, \tag{20}$$

where $\Delta\delta = \Delta\delta_2 - \Delta\delta_1$. Some special cases that make the denominators of equations (19) and (20) being zero should be excluded. One is the case where $\theta_1 = 0$, $\theta_2 = 0$ and $\theta_1 = \pi$, $\theta_2 = \pi$ corresponding to the uniform polarization states on the classical Poincaré sphere, which has been discussed in the main and following texts. The other is that $\theta_3 = \pi$, $\theta_1 = 0$ or $\pi$ and $\theta_3 = 0$, $\theta_2 = 0$ or $\pi$, corresponding to scalar superposition states, which can be detected by using a single centroid ellipse[46]. Additionally, if $\theta_1 = \theta_2 = \theta_3 = \pi/2$, $\varphi_1 = \varphi_2$, the three centroid trajectories may reduce to three straight lines. In this case, $\varphi_m$ ($m = 1, 2$) can be determined by the corresponding orientation angles $\psi_m$, while $\varphi_3$ should be deduced from the length of the correlative orbital line. As the lengths are identical in the cases of $\varphi_3 = k\pi$ and $\varphi_3 = (2-k)\pi$ ($0 \leq k \leq 2\pi$), the time-domain spectra of the x-coordinate variation of two centroid lines 1 and 2 can be used to differentiate these two cases by estimating the phase advance/delay via Fourier transformation, similar to Extended Data Fig. 1.

**Experimental details of detecting and reconstructing photonic 4D spin-orbit states**

In the proof-of-concept experiment, various kinds of 4D spin-orbit states were generated by the combination of the half wave plates, quarter wave plate, and Q plate, as well as the few-mode fibers (Figs. S3 and S4). The optical path differences between the targeted 4D spin-orbit states and reference light were produced by manually pressing the prism in one direction, and meanwhile the dynamical coaxial interference patterns with orthogonal spin components were collected by the high-speed CCD2 (about 180 fps) (see Figs. S3 and S4). The BS3 in Fig. S3 was used to split the 4D spin-orbit states into another path for polarization decomposition via seven procedures by adjusting QWP2 and Pol2 to reconstruct polarization patterns based on the conventional Stokes parameters (Fig. S2a), while another optical path split by BS2 in Fig. S4 was used for two purposes. One is to reconstruct the local polarization distributions of 4D spin-orbit states based on the conventional Stokes polarization decomposition via seven procedures when inserting the box into the optical path and adjusting QWP2 and Pol2 (Fig. S2a). The other is to directly collect the intensity patterns of the same modal states outputted from the FMF by CCD1 when the interference patterns were fastly collected by CCD2 to detect the higher-order modes using our centroid geometric detection. Note that it has not enough time to perform the Stokes polarization decomposition with seven procedures during this process. So we did not provide the polarization patterns reconstructed by the Stokes polarization decomposition in the Supplementary Videos 2 and 3, but just offered the intensity patterns for comparison, see the top left figures in these two Videos.

**Data availability**
Source data are provided with this paper. All other data that support the plots within this paper and other findings of this study are available from the corresponding author upon reasonable request.

**Code availability**
The computer codes that support the plots within this paper and other findings of this study are available from the corresponding author upon reasonable request

**Acknowledgments:** This work was supported by the Natural Science Foundation of China (NSFC) (62275092), the Natural Science Foundation of Hunan Province, China (No. 2025JJ50355 and 2023JJ30114), and the Fundamental Research Funds for the Central Universities.

**Author contributions:** L. Fang conceived the concept and experiments. L. Fang and J. Chen performed the theoretical derivation. J. Chen, J. Cheng, X. Guo, and S. Huang performed the experiments. L. Fang and J. Chen performed data analyses. L. Fang wrote the manuscript and supervised the project. All authors contributed to the discussions and editing of the article.

**Competing interests:** The authors declare no competing interests.

**Additional information**
**Supplementary information** The online version contains supplementary material available at .
**Correspondence and requests for materials** should be addressed to Liang Fang.

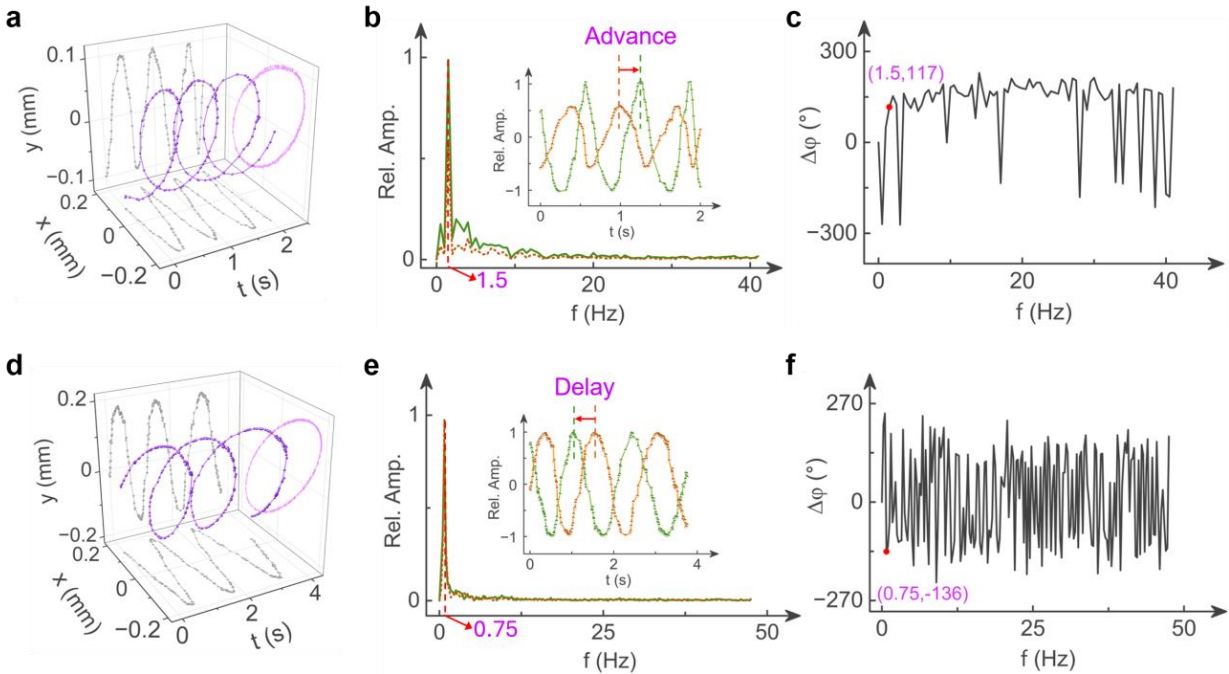

**Extended Data Fig. 1 | Experimental exhibition on how to distinguish the centroid orbiting directions. a** and **d**, Three-dimensional centroid orbiting with different handedness, and their projections onto the $x-t$ and $y-t$ planes. **b** and **e**, The time-domain (insets) and Fourier frequency-domain spectra of x- and y-coordinates of centroid coordinate variation. **c** and **f**, Fourier phase spectra as the differences of frequency-domain signals between x- and y-coordinates, of which the phase differences of peak frequencies can be used for estimating the centroid orbiting directions and thus the signs of orbital ellipticity, i.e., $\Delta\varphi > 0$ or $< 0$ corresponds to the plus or minus signs of ellipticity.

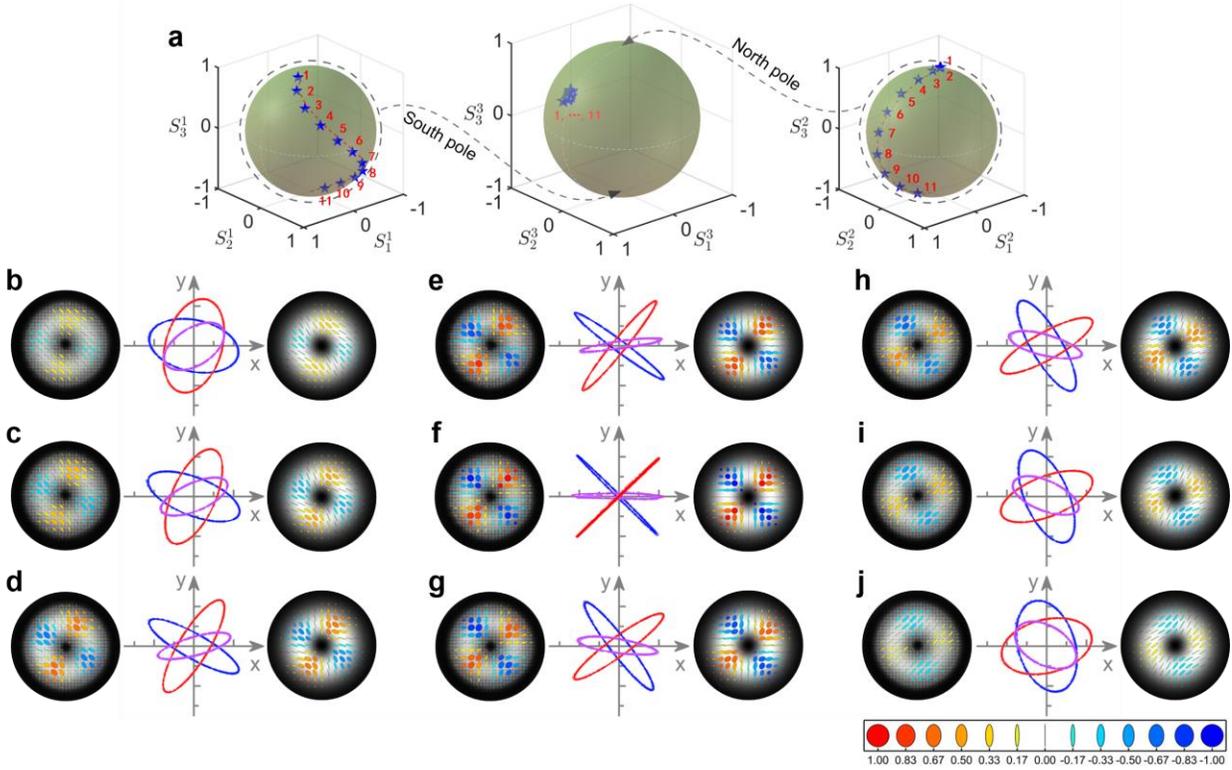

**Extended Data Fig. 2 | Detecting 4D spin-orbit states based on centroid elliptic parameters.** **a**, Experimental results of quantifying 4D spin-orbit states evolved on SU(4) hypersphere, where the states at the south and north poles on middle sphere come from any states on the left and right spheres, respectively. **b-j**, Measured results of 4D spin-orbit states on triple-spheres marked by 2, 3, 4, …, and 10, respectively. The left figures show the measured polarization patterns using the conventional Stokes polarization decomposition via seven procedures. The middle figures show the centroid positions on elliptic orbits with specific radii and azimuthal angles, experimentally extracted from two orthogonally polarized interference patterns. The right figures show the reconstructed patterns based on LG beams with the superposition parameters deduced from the centroid positional information in the middle figures.

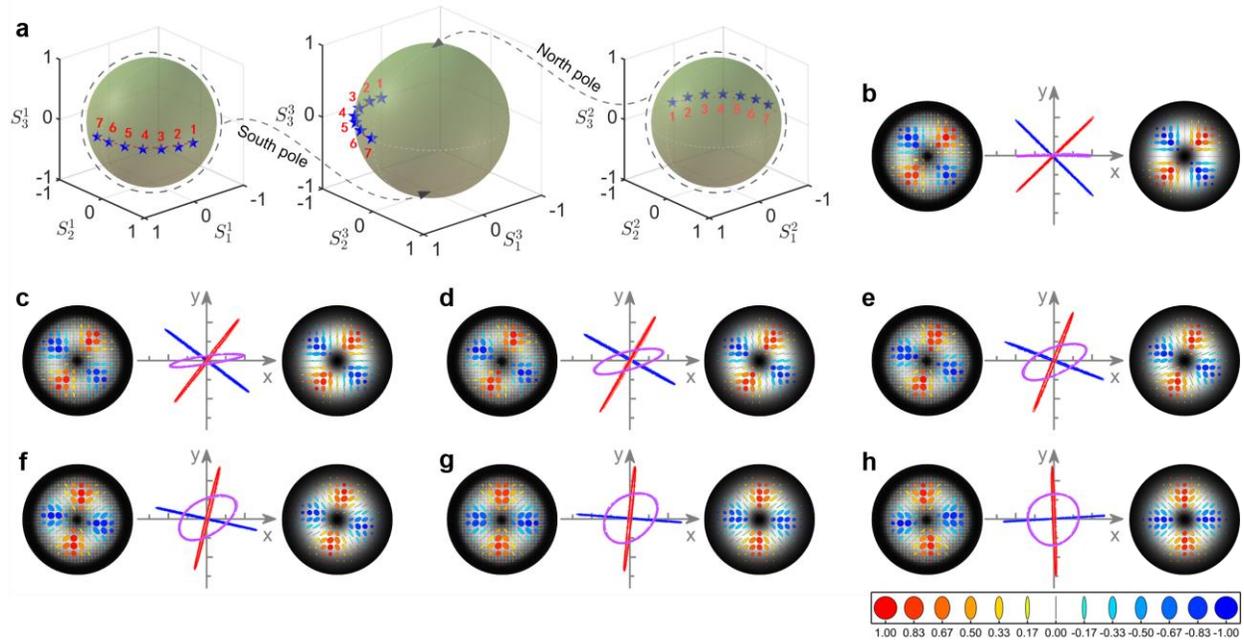

**Extended Data Fig. 3 | Detecting 4D spin-orbit states under the superposition parameters of based on centroid elliptic parameters. a**, Exhibition of quantifying the 4D spin-orbit states evolved along the equators on three Poincaré spheres produced by controlling the QWP behind the Q plate. **b-h**, Experimental results of measuring and reconstructing the 4D spin-orbit states marked with 1, 2,…, 7, respectively, in the trajectory shown on Poincaré sphere. The left figures show the measured patterns using the conventional Stokes polarization decomposition. The middle figures show the extracted centroid orbits (blue and red ellipses) experimentally extracted from two orthogonally polarized interference patterns, and their correlative orbits (magenta ellipses). The right figures show the reconstructed 4D spin-orbit states using LG beams with superposition parameters deduced from the three orbital ellipses.

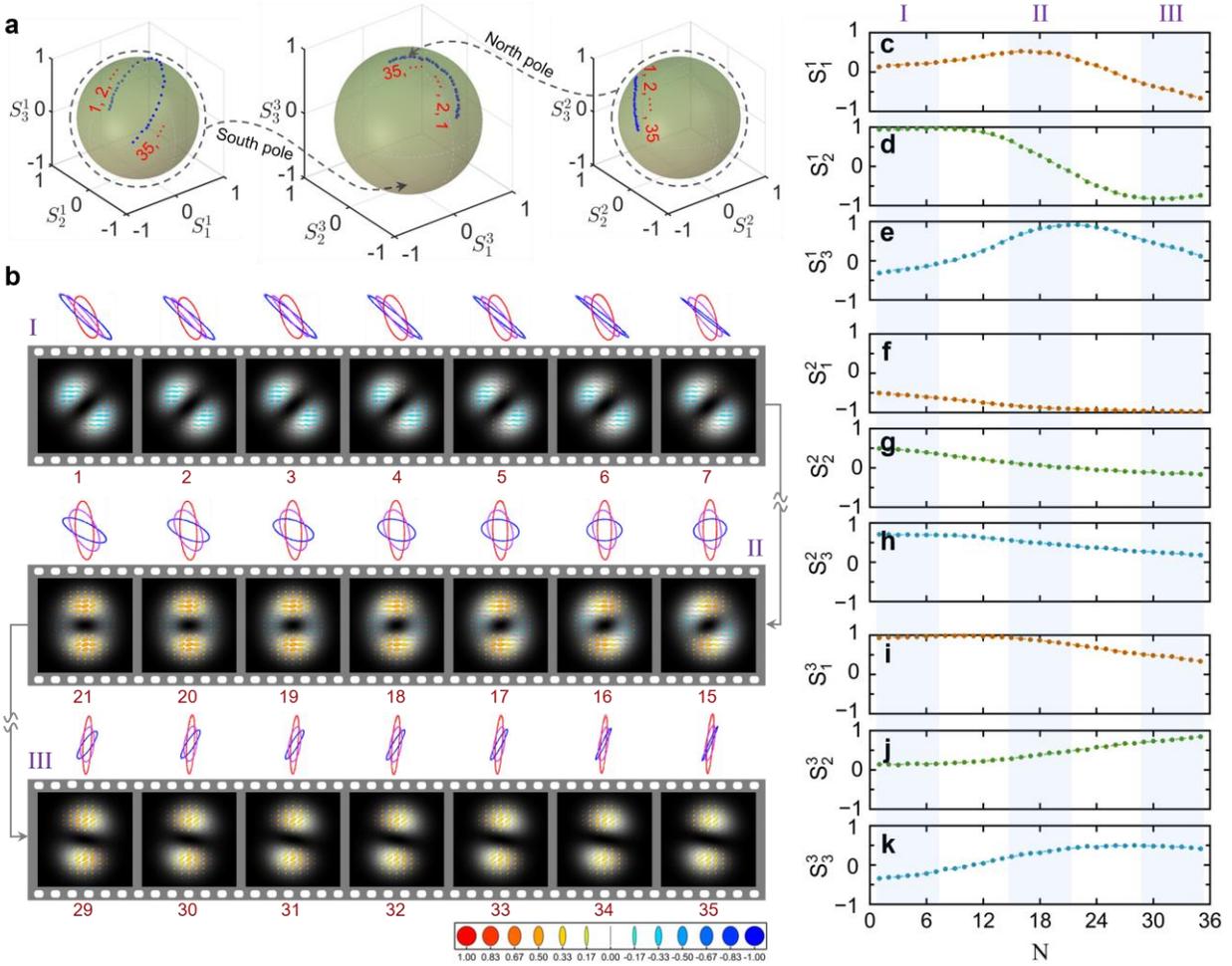

**Extended Data Fig. 4 | Experimental demonstration of quantifying and reconstructing higher-order mode (4D spin-orbit states) evolution along another path produced by twisting the FMF. a**, The higher-order mode evolution is quantified on SU(4) hypersphere. **b**, Reconstructed higher-order modal patterns in the evolution path based on the measured three ellipses shown above. Measured Stokes parameters of higher-order mode evolution on the **c**-**e**, left, **f**-**h** right, and **i**-**k** middle spheres, respectively, where the details of the successive states I, II, and III, as well as beginning and end states, are shown in **b**. The dynamical demonstration and comparison results are demonstrated in Supplementary Video 3.